# Cleavage Tendency of Anisotropic Two Dimensional Materials: ReX$_2$ (X=S, Se) and WTe$_2$

Haifeng Wang[1,2], Erfu Liu[1], Yu Wang[1], Bo Wan[1], Ching-Hwa Ho[3], F. Miao*[1] and X. G. Wan*[1]

[1]National Laboratory of Solid State Microstructures, School of Physics, Collaborative Innovation Center of Advanced Microstructures, Nanjing University, Nanjing 210093, China

[2]Department of Physics, College of Science, Shihezi University, Xinjiang 832003, China

[3]Graduate School of Applied Science and Technology, National Taiwan University of Science and Technology, Taipei 106, Taiwan

**Corresponding Authors**

*E-mail: miao@nju.edu.cn

*E-mail: xgwan@nju.edu.cn

**Author Contributions**

Haifeng Wang, Erfu Liu contributed equally to this work.

**Abstract:**

With unique distorted 1T structure and the associated in-plane anisotropic properties, mono- and few-layer $ReX_2$ (X=S, Se) have recently attracted particular interest. Based on experiment and first-principles calculations, we investigate the fracture behavior of $ReX_2$. We find that the cleaved edges of $ReX_2$ flakes usually form an angle of ~120° or ~60°. In order to understand such phenomenon, we perform comprehensive investigations on the uniaxial tensile stress-strain relation of monolayer and multi-layer $ReX_2$ sheets. Our numerical calculation shows that the particular cleaved edges of $ReX_2$ flakes are caused by unique anisotropic ultimate tensile strengths and critical strains. We also calculate the stress-strain relation of $WTe_2$, which explains why their cleaved edges are not corresponding to the principle axes. Our proposed mechanism about the fracture angle has also been supported by the calculated cleavage energies and surface energies for different edge surfaces.

# Introduction

Two-dimensional transition metal dichalcogenides (2D-TMDs) ($MX_2$, where M denotes a transition metal and X denotes a chalcogen) are promising candidate materials for next-generation flexible optoelectronic applications owing to their mechanical flexibility, chemical and environmental stability, unique optical properties [1-3]. Most of the studied group-VI TMDs exhibit isotropic behaviors [4-6]. Lowering lattice symmetry could induce interesting anisotropic properties. Recently, much research interest has been focused on a new type of 2D-TMDs with low symmetry: rhenium disulfide ($ReS_2$) and rhenium diselenide ($ReSe_2$) (denoted as $ReX_2$) [7-17], both of which exhibit a distorted octahedral phase (denoted as $T_d$). Due to the low lattice symmetry and unique band structures, $ReX_2$ show pronounced in-plane anisotropic properties with possible device applications. For example, few-layer $ReX_2$ sheets have been demonstrated as channel materials for field effect transistors [8,9,11], integrated digital inverters [9] and photodetectors [11,16,17]. Similar to $ReX_2$, $WTe_2$ forms low-symmetry distorted 1T structure [18] while exhibiting anisotropic electronic, optical, vibrational and thermal dynamical properties [18-25].

Mechanical exfoliation from bulk layered crystals has been a common approach to obtain atomically thin 2D materials, which usually show tendency to fracture along certain in-plane crystallography orientations, especially for those with low symmetry. For example, the morphology of the exfoliated thin $ReS_2$/$ReSe_2$ flakes was found to exhibit a quadrilateral shape with inner angles of ~60° or ~120° [9,10,16,17], which was suggested to be caused by the weakest breaking strength along two principle axes [9]. Despite the similar lattice structure, $WTe_2$ thin flakes are not preferentially cleaved along their crystal axes [20,23,24]. To our best knowledge, no theoretical study about such phenomenon has been performed.

In this article, we present a detailed study of cleavage tendency of mono- and multi-layer $ReS_2$, $ReSe_2$ and $WTe_2$. Our experiments confirm that the cleaved edges of $ReS_2$ thin flakes indeed tend to form the angle of 120º or 60º, with major results shown in Figure 1. Our first-principles calculations successfully explain such experimental results. Our numerical calculations further reveal the reason why the edges of cleaved WTe$_2$ thin flakes are not related with their principle axes. Our study suggests that the cleaved edges of 2D layered materials are determined by the tensile stress.

## Method

Single crystals of ReS$_2$ were grown by the same Br$_2$-assisted chemical vapor transport method described in Ref. [7]. We used a standard mechanical exfoliation method to isolate mono- and few-layer ReS$_2$ flakes. Similarly, the WTe$_2$ thin films were mechanical exfoliated from single crystals (HQ-graphene, Inc.) onto the silicon substrate covered by 285 nm SiO$_2$.

The first-principle calculations were carried out with the Vienna *ab-initio* Simulation Package (VASP) [26] based on density functional theory (DFT). The Perdew-Burke-Ernzerh of (PBE) exchange-correlation functional [27] along with the projector-augmented wave (PAW) potentials was employed for the self-consistent total energy calculations and geometry optimization. The kinetic energy cutoff for the plane wave basis sets was chosen to be 550 eV for ReX$_2$ and 500 eV for WTe$_2$. The Brillioun zone was sampled using $5\times5\times1$ (for mono- and multi-layer ReX$_2$) and $5\times5\times5$ (for bulk ReX$_2$) Monkhorst-Pack k-point grid. For WTe$_2$, the Monkhorst-Pack k-point grids are $14\times7\times1$ (for mono- and multi-layer WTe$_2$) and $14\times7\times3$ (for bulk WTe$_2$). Atomic positions were relaxed until the energy differences were converged within $10^{-5}$ eV and the maximum Hellmann-Feynman force on any atom was less than 0.01 eV/Å. Based on the optimized lattice structure, we performed further calculation. Since van der Waals (vdW) interaction is important in layered materials [28-31], by using a semi-empirical DFT-D$_2$ method [32], we considered the vdW correction in all calculations except the stress-strain relations of the monolayer ReX$_2$ and WTe$_2$.

The theoretical stress-strain relation was calculated by using a standard method [33], which requires rectangular unit cell to impose uniaxial tensions. To calculate the stress-strain relations, we applied a series of incremental uniaxial tensile strains along one direction of the supercell and relaxed the lattice along the orthogonal direction. Each of the corresponding conjugate stress components was less than 0.05 GPa. A vacuum of 20 Å along **c** direction (inter-layer direction) was included to avoid the interaction between the periodically repeated structures.

In 2D system, stress calculated by DFT has to be modified to avoid the force being averaged over the entire simulation cell including vacuum space [33-35]. In order to compare with experimental results directly, the stress was rescaled by $Z/d_0$ to obtain equivalent stress, where Z is the cell length in **c** direction and $d_0$ is the effective thickness of the system [33-35]. Here, the effective thickness was taken to be 0.70 nm for $ReS_2$ [7], 0.66 nm for $ReSe_2$ [12], 0.70 nm for $WTe_2$ [18] respectively.

## Results

As shown by the top view of monolayer $ReS_2$ crystal structure in Figure 2(a), $ReX_2$ crystallize in a distorted 1T structure [7]. The clusters of $Re_4$ units are arranged in a diamond-like shape, and form a quasi-one-dimensional chain inside each monolayer [7]. There are two principle crystal axes, the **b** and **a** axes, with an angle about 120º formed between them. The **b**-axis corresponds to the direction of Re-Re atomic chain as shown in Figure 2. Compared to those metal atoms in those widely studied group-VI TMDs, such as $MoS_2$, the extra one valence electron in each rhenium atom leads to the formation of the Re-Re bond in $ReX_2$. Therefore, the forming of $Re_4$ cluster significantly affects the interaction between Re-Re dimers [7]. The mechanical properties of $ReX_2$ strongly depend on such strongly covalent-bonded Re clusters. Regarding the $Re_4$ diamond-shape clusters, there are four most-notable directions: **a**, **b** lattice vectors and two diagonal directions of unit cell. Specifically, as shown in Figure 2(a), they are Re1-Re2 direction (0°, i.e., **b**-axis), Re1-Re4 direction (~120°, i.e., **a**-axis), Re1-Re3 direction (~60°, i.e., one of the diagonal direction of $Re_4$ diamond), and Re2-Re4 direction (~150°, i.e., the other diagonal direction of $Re_4$ diamond).

In order to study the breaking strengths along these four typical directions, we adopted three 24-atom orthogonal supercells: 0°-90°, 30°-120° and 60°-150° supercells as shown in Figure 2(b)-(d) respectively. The calculated stress-strain relations of monolayer ReS$_2$ and ReSe$_2$ along six special directions are presented in Figure 3 (including the aforementioned four typical directions adding four perpendicular directions; here 60° and 90° directions are perpendicular to each other, thus there are total six special directions: 0°, 30°, 60°, 90°, 120°, 150°). We also calculated the stress-strain relations of bilayer ReX$_2$, no obvious difference was found compared to monolayer ReX$_2$ except the ultimate stresses are slightly lower for the bilayer case. We also calculated the stress-strain relation for trilayer ReX$_2$, the stress-strain behavior of which was found to be similar to that of bilayer ReX$_2$. Since vdW interaction usually plays important role in interlayer coupling, we also compared the results of multi-layer ReX$_2$ with and without vdW interaction considered. Our numerical results show that vdW interaction only slightly affects the results, with summarized results listed in Table 1. We hence focused on the stress-strain relation of monolayer ReX$_2$.

By fitting the initial stress-strain curves on the linear region up to 2% strain, we estimated the elastic properties of monolayer ReX$_2$. ReX$_2$ show nearly isotropic in-plane elastic response with Young's modulus $E_{\text{Re}S_2} \approx 190$ GPa, $E_{\text{Re}Se_2} \approx 140$ GPa and Poisson's ratio $\mu_{\text{Re}S_2} \approx 0.225$, $\mu_{\text{Re}Se_2} = 0.223$. It is in sharp contrast to black phosphorene (with similar strong in-plane anisotropic structure), which show obvious in-plane anisotropic elastic responses [34]. The Young's modulus and ultimate strengths of ReX$_2$ are also much lower than that of MoX$_2$ (X=S, Se) [35,36]. When the applied strain increases, the calculated stress-strain behaviors of ReX$_2$ monolayer become nonlinear. The specific ultimate tensile strengths and corresponding critical strains along six different directions are listed in Table 1. ReS$_2$ and ReSe$_2$ show similar strain-stress relations with strong orientation-dependence. The critical strains of monolayer ReS$_2$ and ReSe$_2$ are also close to each other (shown in Table 1). Meanwhile, the ultimate stresses of ReS$_2$ are considerable stronger than those of ReSe$_2$, since the bond of Re-S is much tighter than that of Re-Se.

While the shortest Re-Re distance in Re-Re chains is along **b**-axis (0° direction), the ultimate strengths and critical strains are the highest along 150° direction in ReX$_2$. Compared with other directions except 150°, the 0° direction shows apparently larger ultimate strengths. Cleavage tendency is usually determined by lower strengths along certain directions. As shown in Table 1, for both materials, the ultimate strengths along 30° and 90° directions are much lower compared to other directions. The lowest ultimate strength appears at 30° direction, and the ultimate strengths of 90° direction are only slightly larger than those of 30° direction. Therefore, during the exfoliation, ReS$_2$ and ReSe$_2$ flakes have larger probabilities to break along the directions of 30° and 90°. In most cases, thin flakes may break nearly simultaneously along these two directions. Since the direction of 30° is perpendicular to **a**-axis, and the direction of 90° is perpendicular to **b**-axis, ReS$_2$ and ReSe$_2$ thin flakes preferentially crack along these two axes, resulting in the angles of 120° and 60° appearing with the largest probability. Such theoretical results are fully consistent with the experimental measurement as shown in Figure 1.

Although the ultimate strengths along 30° and 90° directions are almost the same for both ReS$_2$ and ReSe$_2$ flakes, the smaller critical strain makes the direction of 90° slightly easier to crack. Therefore, the probability of the presence of **b**-axis is larger than that of **a**-axis. This is also consistent with the previously reported observations. For example, Y. -C. Lin *et al.* [13] found that during the exfoliation, the edges of bilayer ReS$_2$ flakes were always oriented along the Re-Re chains, i.e., **b**-axis. D. Chenet *et al.* [14] examined the cleaved edges of mono- and few-layer ReS$_2$ samples by using Raman spectroscopy, and confirmed that in many cases the edges were parallel to **b**-axis. L. Hart *et al.* [15] also found similar phenomenon that the crystallographic **b**-axis of ReS$_2$ flakes frequently formed the longer edge of cleaved crystals, clearly indicating larger appearance probability for **b**-axis than **a**-axis. This study is in good agreement with our experiment results. As shown in Figure 1, the typical ReS$_2$ flake shows not only the appearance of angle ~120° but also the longer edges of **b**-axis. The difference between the critical strains of 30° and 90° directions may be induced by the space difference between two vicinal diamond shaped clusters in **b** and **a** directions. As shown in Figure 2(b)-(c), the vicinal Re$_4$ clusters are separated by 0.30 nm along the 30° direction, while the distance is 0.35 nm along the 90° direction.

By comparing to ReX$_2$, WTe$_2$ has shown many similarities, including the same distorted 1T lattice structure, similar anisotropic electronic, optical, vibrational and thermal dynamical properties [8,9,11,16-25]. Both materials exhibit metal atom chains along one of the principle axes as shown in Figure 2 and Figure 4. However, it seems that the cleavage edges of WTe$_2$ flakes are generally not related with the principle axes [20,23,24]. We thus performed similar calculations to explore the physics behind such difference.

By fitting the initial linear stress-strain curves in small strain regime, the mechanical properties of monolayer WTe$_2$ were obtained with results shown in Figure 4(c). Unlike monolayer ReX$_2$, the elastic response of monolayer WTe$_2$ shows obvious anisotropy. We calculated Young's modulus along two principle axes, with values of $E_X = 117$ GPa and $E_Y = 140$ GPa, which implies that the W-Te bonds along the Y-axis may be stronger than those along the X-axis. The Possion's ratios along two principle axes are also highly anisotropic, with values of $\mu_X \approx 0.25$ and $\mu_Y \approx 0.35$ along the X- and Y-axis, respectively. The anisotropy of Poisson's ratios indicate that monolayer WTe$_2$ is less responsive under strain along the W-W chain direction than in the perpendicular direction.

Then we focused on the strain effects of monolayer WTe$_2$ under large strains. The calculated maximum Cauchy stress for uniaxial tension along the X-axis is 9.35 GPa, at critical strain $\varepsilon_X = 0.12$. While along the Y-axis, WTe$_2$ was found to be much stronger with ultimate strength of 14.96 GPa. Monolayer WTe$_2$ shows superior flexibility along Y-axis with a large critical strain $\varepsilon_Y = 0.16$. The differences of the ultimate strengths and critical strains between two principle axes are large, which implies that the two axes usually do not crack simultaneously during exfoliation. X-axis should be much easier to crack because of the much smaller ultimate strength and critical strain compared to Y-axis. Thus in contrast with ReX$_2$, WTe$_2$ flakes usually do not form shapes with inner angle of 90°. Based on these numerical results, one can expect that WTe$_2$ thin flakes may show long-strip shape along the Y-axis. Such results agree with the prior experimental works [23,24]. We also found that the exfoliated WTe$_2$ thin flakes naturally formed long-strip shape, as shown in Figure 4(d).

To further study the cleavage tendency of ReX$_2$ and WTe$_2$, we also calculated the cleavage energy E$_{cl}$ for different edge surfaces. E$_{cl}$ is defined as the minimum energy required to overcome interlayer force during exfoliation processes [37,38]. The structure of the bulk ReS$_2$ is shown in Figure 5(a), where the side surfaces {100} and {010} represent the surfaces containing the 0° and 60° directions, respectively. Other edge surfaces are similar and not shown. As an example, the calculated E$_{cl}$ is shown in Figure 5(b) by using edge surface {100} of ReS$_2$. Firstly, we constructed a four-layer slab serving as the model of bulk, where three layers were fixed and the other monolayer was flexible as an exfoliated layer. The separation distance in the equilibrium configuration is defined as zero. A vacuum layer (15Å) is incorporated into the four-layer slab to avoid artificial interactions between two neighboring slabs. As shown in Figure 5(b), cleavage energy increases with the distance between the exfoliated 1L and bulk, and reaches convergence at about 2.5Å. The cleavage energy of edge surface {100} is about 2.15 J/m$^2$. The theoretical cleavage strength curve was obtained by taking the derivative of E$_{cl}$ with respect to distance. The obtained value is about 20.91 GPa, which is larger than the ultimate tensile strength of this direction (~15.21GPa).

In Table 2, the calculated cleavage energies and strengths for all typical edge surfaces of ReX$_2$ and WTe$_2$ are listed with corresponding ultimate tensile strengths for comparison. Among all edge surfaces of ReX$_2$, the edge surface containing 0° has the lowest cleavage energy and cleavage strength. The edge surface containing 120° has the second lowest cleavage energy and cleavage strength. Thus, these two edge surfaces have the largest probabilities to appear, which agrees with the results on the tensile stress-strain relation. Similar results were also found for WTe$_2$: the {100} surface has not only smaller cleavage energy and cleavage strength, but also smaller ultimate tensile strength.

Another important factor related to the specific edge termination of a certain crystal is its surface energy. Surface energy is the energy required to create new surfaces when crystals are peeled [39,40]. According to the Gibbs-Curie-Wulff law, surface with lower surface energy is the preferable lateral face of a material. We calculated and compared the surface energies of all typical terminations for $ReX_2$ and $WTe_2$, with results shown in Table 2. The surface energy of $ReX_2$ is defined by the following equation:

$$\varepsilon = \frac{E(nReX_2) - nE(ReX_2)}{2A}$$

where $E(nReX_2)$ is the total energy of $ReX_2$ surface slab, $n$ is the number of the atoms in the surface slab, $E(ReX_2)$ is the bulk energy per atom and $A$ is the surface area. As expected, the surface containing 0° has the lowest energy (0.78 J/m$^2$ for $ReS_2$ and 0.72 J/m$^2$ for $ReSe_2$), followed by the surface containing 120° (0.88 J/m$^2$ for $ReS_2$ and 0.86 J/m$^2$ for $ReSe_2$). For $WTe_2$, the surface energy of {100} surface (0.77 J/m$^2$) is considerably lower than {010} surface (1.28 J/m$^2$), which further confirmed the above results that the X-axis of $WTe_2$ flakes is much easier to crack.

## Summary

In conclusion, based on the experimental observation and first-principle calculations, we investigated the cleavage tendency of three types of $T_d$-TMDs, $ReX_2$ (X=S, Se) and $WTe_2$. The ultimate tensile strengths and critical strains of these three low-symmetry materials were found to be strongly orientation-dependent. We found the ultimate stresses of $ReX_2$ flakes along the directions perpendicular to **a** and **b** lattice vectors are much lower than other directions, which is responsible to the observed phenomenon that cleavage edges preferentially lie along their crystal axes. Additionally, along the direction perpendicular to **b**-axis, our calculations shew that the critical strain are significantly lower than other directions. This explains why the **b**-axis usually forms the longer edge of the cleaved flakes. For $WTe_2$, two principle axes have quite different ultimate strengths and critical strains, which explains why WTe$_2$ tends to form long-strip shape. These conclusions were further confirmed by the comparison of cleavage energies and surface energies for different edge surfaces.

**ACKNOWLEDGMENTS**

The work was supported by National Key R&D Program of China (No. 2017YFA0303203), the National Key Basic Research Program of China (2015CB921600, 2013CBA01603), National Natural Science Foundation of China (11525417, 11374137, 11374142, 61574076, 61625402, 51721001), the Natural Science Foundation of Jiangsu Province (BK20140017, BK20150055), Doctoral Fund of Ministry of Education of China (Grant No. 20130091110003), and Fundamental Research Funds for the Central Universities and the Collaborative Innovation Center of Advanced Microstructures.

**Reference**

[1] Q. H. Wang, K. Kalantar-Zadeh, A. Kis, J. N. Coleman, and M. S. Strano, Nat. Nanotech. **7,** 699 (2012).

[2] B. Radisavljevic, A. Radenovic, J. Brivio, V. Giacometti, and A. Kis, Nat. Nanotech. **6,** 147 (2012).

[3] M. Chhowalla, H. S. Shin, G. Eda, L. J. Li, and K. P. Loh, Nat. Chem. **5,** 263 (2013).

[4] A. K. Geim, and I. V. Grigorieva, Nature, **499**, 419 (2013).

[5] Q. H. Wang, K. Kalantar-Zadeh, A. Kis, J. N. Coleman, and M. S. Strano, Nat. Nanotech. **7,** 699 (2012).

[6] K. F. Mak, C. G. Lee, J. Hone, J. Shan, and T. F. Heinzet, Phys. Rev. Lett. **105,** 136805 (2010).

[7] S. Tongay, H. Sahin, C. Ko, A. Luce, W. Fan, K. Liu, J. Zhou, Y. -S. Huang, C. -H. Ho, J. Y. Yan, D. F. Ogletree, S. Aloni, J. Ji, S. S. Li, J. B. Li,.; F. M. Peeters, and J. Q. Wu, Nat. Commun. **5**, 4252 (2014).

[8] C. M. Corbet, C. McClellan, A. Rai, S. S. Sonde, E. Tutuc, and S. K. Banerjee, ACS Nano **9,** 363 (2015).

[9] E. Liu, Y. Fu, Y. Wang, Y. Feng, H. Liu, X. Wan, W. Zhou, B. Wang, L. Shao, C. -H. Ho, et al, Nat. Commun. **6,** 6991 (2015).

[10] Y. Feng, W. Zhou, Y. Wang, J. Zhou, E. Liu, Y. Fu, Z. Ni, X. Wu, H. Yuan, F. Miao, B. Wang, X. G. Wan, and D. Y. Xing, Phys. Rev. B **92,** 054110 (2015).

[11] E. Zhang, Y. Jin, X. Yuan, W. Wang, C. Zhang, L. Tang, S. Liu, P. Zhou, W. Hu, and F. Xiu, Adv. Funct. Mater. **25,** 4076 (2015).

[12] D. Wolverson, S. Crampin, A. S. Kazemi, A. Ilie, and S. J. Bending, ACS Nano **8,** 11154 (2014).

[13] Y. -C. Lin, C. -H. Yeh, T. Björkman, Z. -Y. Liang, C. -H. Ho, Y. -S. Huang, P. -W. Chiu, A. V. Krasheninnikov, and K. Suenaga, ACS Nano **9,** 11249 (2015).

[14] D. Chenet, O. Aslan, P. Huang, C. Fan, A. Zande, T. Heinz, and J. Hone, Nano Lett. **15,** 5667 (2015).

[15] L. Hart, S. Dale, S. Hoye, L. W. James, and D. Wolverson, Nano Lett. **16,** 1381

(2016).

[16] E. Zhang, P. Wang, Z. Li, H. F. Wang, C. Y. Song, C. Huang, Z. -G. Chen, L. Yang, K. T. Zhang, S. H. Lu, W. Y. Wang, S. S. Liu, H. H. Fang, X. H. Zhou, H. G. Yan, J. Zou, X. G. Wan, W. D. Hu, P. Zhou, and F. X. Xiu, ACS Nano, **10,** 8067 (2016).

[17] E. Liu, M. Long, J. Zeng, W. Luo, Y. Wang, Y. Pan, W. Zhou, B. Wang, W. Hu, Z. Ni, Y. You, X. Zhang, S. Qin, Y. Shi, K. Watanabe, T. Taniguchi, H. Yuan, H. Hwang, Y. Cui, F. Miao, and D. Xing, Adv. Funct. Mater. **26,** 1938 (2016).

[18] M. N. Ali, J. Xiong, S. Flynn, J. Tao, Q. D. Gibson, L. M. Liang, T. Liang, N. Haldolaarachchige, M. Hirshberger, N. P. Ong, and R. J. Large, Nature **514,** 205 (2014).

[19] A. A. Soluyanov, D. Gresch, Z. J. Wang, Q. S. Wu, M. Troyer, X. Dai, and B. A. Bernevig, Nature **527,** 495 (2015).

[20] X. C. Pan, X. Chen, H. Liu, Y. Feng, F. Song, X. G. Wan, Y. Zhou, Z. Chi, Z. Yang, B. Wang, Y. Zhang, and G. Wang, Nat. Commun. **6,** 7805 (2015).

[21] I. Pletikosic, M. N. Ali, A.V. Fedorov, R. J. Cava, and T. Valla, Phys. Rev. Lett. **113,** 216601 (2014).

[22] C. C. Homes, M. N. Ali, and R. J. Cava, Phys. Rev. B **92,** 161109 (2015).

[23] Q. J. Song, X. C. Pan, H. F. Wang, K. Zhang, Q. H. Tan, P. Li, Yi. Wan, Y. L. Wang, X. L. Xu, M. L. Lin, X. G. Wan, F. Q. Song, and L. Dai, Sci. Rep. **6,** 29254 (2016).

[24] Y. Wang, E. Liu, H. liu, Y. Pan, H. Zhang, J. Zeng, Y. Fu, M. Wang, K. Xu, Z. Huang, Z. Wang, H. Lu, D. Xing, B. Wang, X. Wan, and F. Miao, Nat. Commun. **7,** 13142 (2016).

[25] G. Liu, H. Y. Sun, J. Zhou, Q. F. Li, and X. G. Wan, New J. Phys. **18,** 033017 (2016).

[26] G. Kresse, and J. Furthmuller, Phys. Rev. B **54,** 11169 (1996).

[27] J. P. Perdew, K.Burke, and M. Ernzerhof, Phys. Rev. Lett. **77,** 3865 (1996).

[28] Z. Zheng, X. Wang, and W. Mi, J. Phys.: Condens. Matter **28**, 505003 (2016.)

[29] M. Zhao, W. Zhang, Ma. Liu, C. Zou, K. Yang, Y. Yang, Y. Dong, L. Zhang, and

S. Huang, Nano Res. **9,** 3772 (2016).

[30] Q. Zhang, W. Wang, X. Kong, R. G. Mendes, L. Fang, Y. Xue, Y. Xiao, M. H. Rümmeli, S. Chen, and L. Fu, J. Am. Chem. Soc. **138,** 11101 (2016).

[31] K. Dileep, R. Sahu, S. Sarkar, S. C. Peter, and R. Datta, J. Appl. Phys. **119,** 114309 (2016).

[32] S. Grimme, J. Comput. Chem. **27,** 1787 (2006).

[33] T. Li, J. W. Morris, and D. C. Chrzan, Phys. Rev. B **70,** 054107 (2004).

[34] Q. Wei, and X. Peng, Appl. Phys. Lett. **104,** 251915 (2014).

[35] T. Li, Phys. Rev. B **85,** 235407 (2012).

[36] J. Li, N. V. Medhekar, and V. B. Shenoy, J. Phys. Chem. C **117,** 15842 (2013).

[37] P. Lazar, and R. Podloucky, Phys. Rev. B **75,** 024122 (2007).

[38] P. Lazar, and R. Podloucky, Phys. Rev. B **78,** 104114 (2008).

[39] P. Raybaud, J. Hafner, G. Kresse, and H. Toulhoat, Surf. Sci. **407**, 237 (1998).

[40] H. Schweiger, P. Raybaud, G. Kresse, and H. Toulhoat, J. Catal. **207,** 76 (2002).

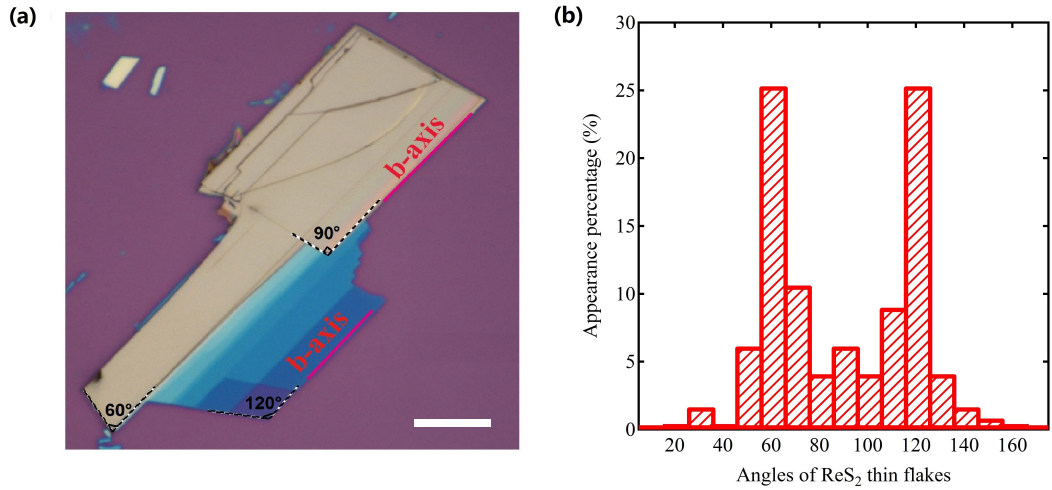

FIG. 1. (a) Optical image of a typical ReS$_2$ thin flakes, which shows not only the most common angle of 60° (or 120°) but also the well-defined longer edges along the **b**-axis. The scale bar is 10 μm. (b) The statistics of inner angles for over 50 ReS$_2$ thin flakes, showing the larger appearance percentage for 60° and 120°.

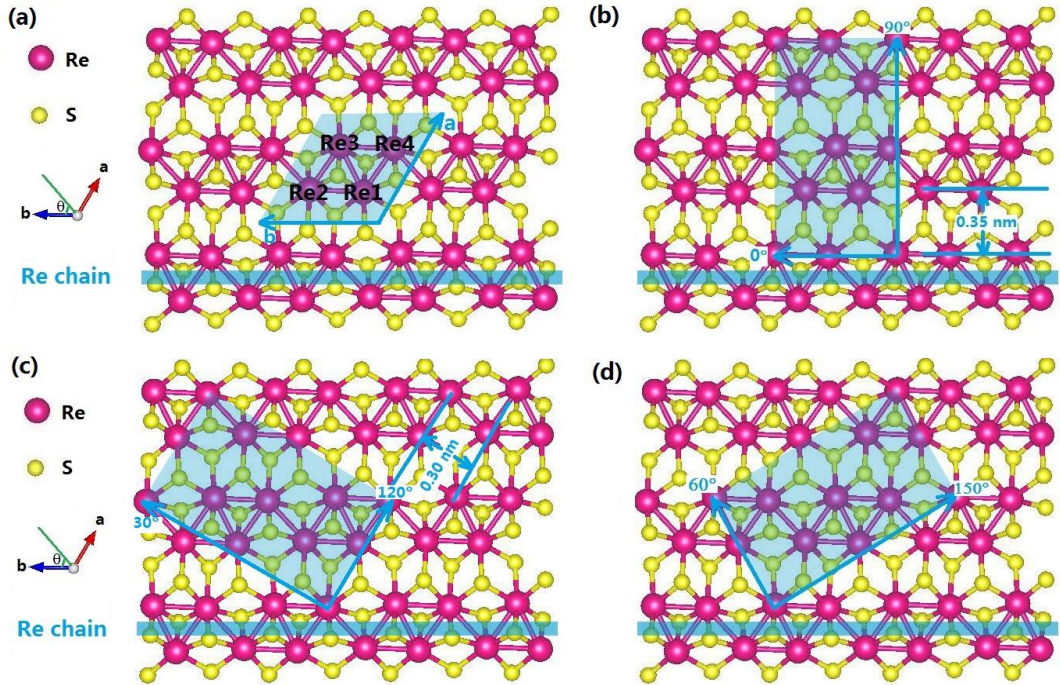

FIG. 2. (a) Atomic configuration of monolayer ReS$_2$ with distorted 1T structure. The parallelepiped is unit cell (blue area) which contains four rhenium atoms and eight sulfur atoms. The angle between **a** and **b**-axis of the unit cell is 119.1°. Rhenium chain along **b**-axis is set as 0° direction and other directions are defined by the angle (θ) from the rhenium chain. (b)-(d) The 24-atoms orthorhombic supercells (blue area) are used to perform stress-strain relation calculations under uniaxial tension, (b) 0°-90° supercell, (c) 30°-120° supercell and (d) 60°-150° supercell.

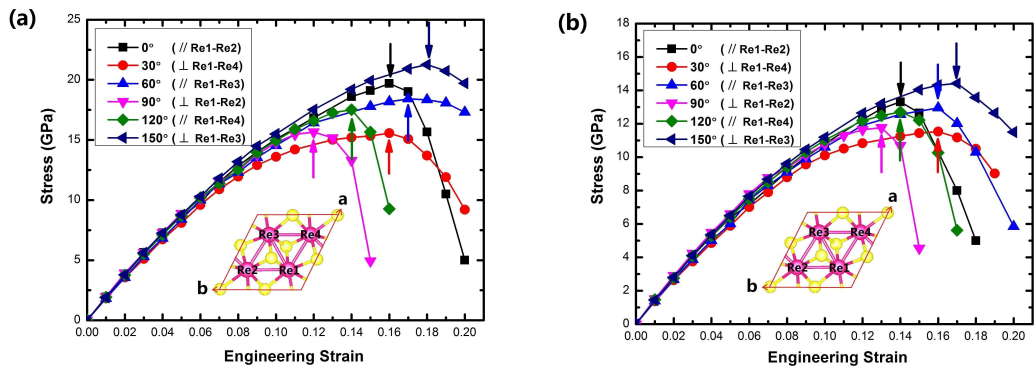

FIG. 3. Calculated stress-strain relations of monolayer (a) ReS$_2$ and (b) ReSe$_2$ under uniaxial tensions along six typical directions. The largest stress along a certain direction is marked by vertical arrow.

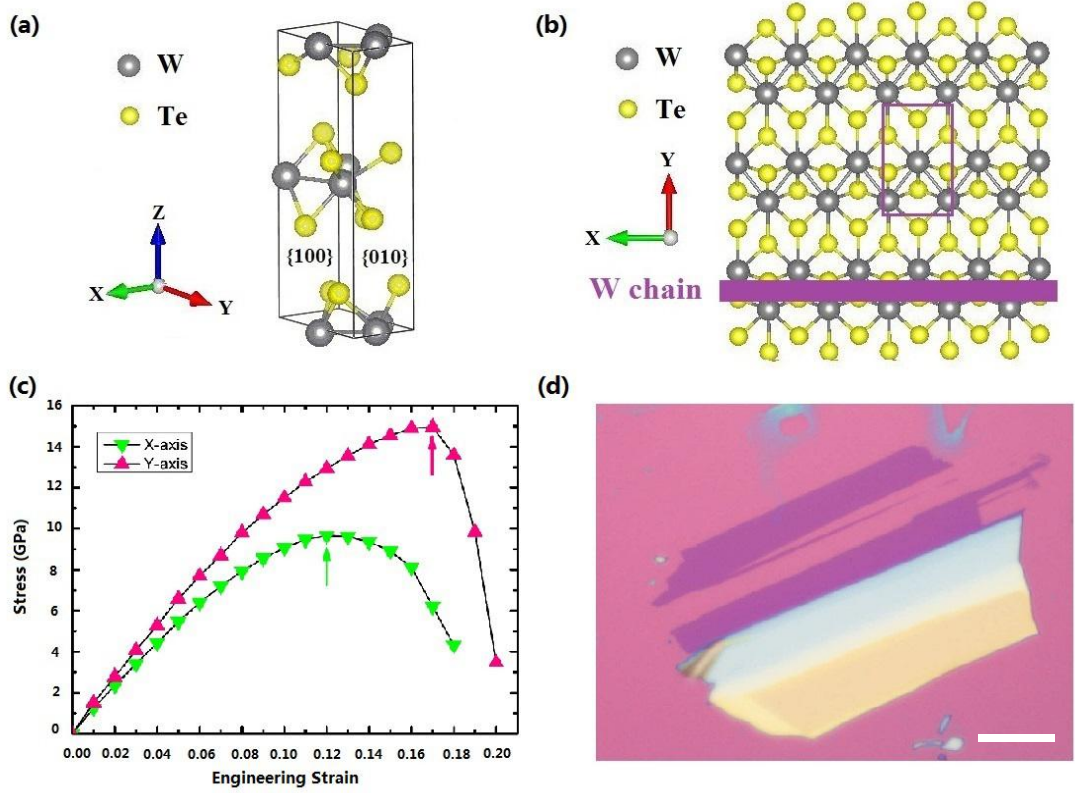

FIG. 4. (a-b) Optimized geometric structure of bulk and monolayer WTe$_2$. The rectangular unit cell is used in stress-strain calculations. (c) Calculated stress-strain relations of monolayer WTe$_2$ under uniaxial tensions along two principle axes. The largest stress along a certain direction is marked by vertical arrow. (d) Optical image of a typical WTe$_2$ thin flake with long-strip shape. The scale bar is 10 μm.

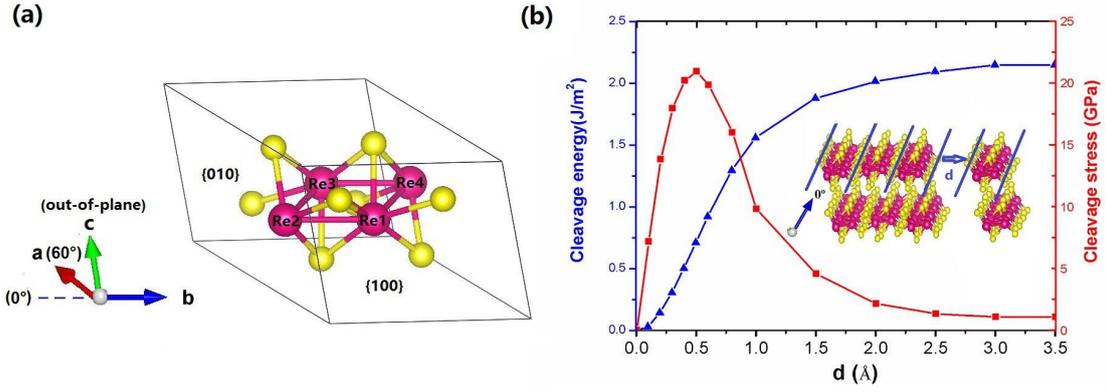

FIG. 5. (a) Optimized geometric structure of bulk ReS$_2$, the {100} surface is edge surface containing 0° (**b**-axis), and the {010} surface is edge surface containing 60°. (b) Cleavage energy E$_{cl}$ in J/m$^2$ (blue line) and cleavage strength in GPa (red line) as a function of separation distance d for the edge surface of ReS$_2$ containing 0° (**b**-axis). Inset: separating a monolayer from its neighboring trilayer.

Table 1. Calculated ultimate tensile strengths and critical strains of ReS$_2$ and ReSe$_2$ under six typical directions and WTe$_2$ under two axes.

| Direction | | Ultimate Strength σ (GPa) | Critical strain ε | Ultimate Strength σ (GPa) | | Critical strain ε | |
|---|---|---|---|---|---|---|---|
| | | Monolayer | | Multi-layer | | | |
| | | | | with vdW | no vdW | with vdW | no vdW |
| ReS$_2$ | 0° (parallel to Re1-Re2) | 19.69 | 0.16 | 19.16 | 19.25 | 0.16 | 0.16 |
| | 30° (perpendicular to Re1-Re4) | 15.56 | 0.16 | 15.15 | 15.24 | 0.16 | 0.15 |
| | 60° (parallel to Re1-Re3) | 18.41 | 0.17 | 17.96 | 18.10 | 0.17 | 0.16 |
| | 90° perpendicular to Re1-Re2 | 15.66 | 0.12 | 15.21 | 15.52 | 0.12 | 0.11 |
| | 120° (parallel to Re1-Re4) | 17.51 | 0.14 | 17.06 | 17.13 | 0.14 | 0.13 |
| | 150° perpendicular to Re1-Re3 | 21.26 | 0.18 | 20.96 | 21.05 | 0.17 | 0.17 |
| ReSe$_2$ | 0° (parallel to Re1-Re2) | 13.31 | 0.14 | 13.01 | 13.24 | 0.14 | 0.13 |
| | 30° (perpendicular to Re1-Re4) | 11.52 | 0.16 | 11.36 | 11.44 | 0.15 | 0.16 |
| | 60° (parallel to Re1-Re3) | 12.85 | 0.16 | 12.01 | 12.55 | 0.16 | 0.15 |
| | 90° (perpendicular to Re1-Re2) | 11.76 | 0.13 | 11.05 | 11.23 | 0.12 | 0.13 |
| | 120° (parallel to Re1-Re4) | 12.70 | 0.14 | 12.10 | 12.51 | 0.14 | 0.14 |
| | 150° (perpendicular to Re1-Re3) | 14.41 | 0.17 | 13.98 | 14.21 | 0.16 | 0.17 |
| $WTe_2$ | X-axis | 9.35 | 0.12 | 8.00 | 8.10 | 0.12 | 0.11 |
| | Y-axis | 14.96 | 0.16 | 12.85 | 12.26 | 0.15 | 0.15 |

Table 2. Calculated cleavage energies，cleavage strengths，ultimate tensile strengths (multi-layer with vdW correction) and surface energies of $ReS_2$, $ReSe_2$ and $WTe_2$ for typical edge surfaces.

| | Surface | Cleavage energy (J/m$^2$) | Cleavage strength (GPa) | Ultimate tensile stress (GPa) | Surface energy (J/m$^2$) |
|---|---|---|---|---|---|
| $ReS_2$ | containing 0° | 2.15 | 20.91 | 15.21 | 0.78 |
| | containing 30° | 4.06 | 29.80 | 17.06 | 1.20 |
| | containing 60° | 4.12 | 28.84 | 20.96 | 1.95 |
| | containing 90° | 3.42 | 25.46 | 19.16 | 1.17 |
| | containing 120° | 2.68 | 22.02 | 15.15 | 0.88 |
| | containing 150° | 2.88 | 23.26 | 17.96 | 1.06 |
| $ReSe_2$ | containing 0° | 1.69 | 15.22 | 11.05 | 0.72 |
| | containing 30° | 2.96 | 22.43 | 12.10 | 1.18 |
| | containing 60° | 3.29 | 21.76 | 13.98 | 1.71 |
| | containing 90° | 2.72 | 19.21 | 13.01 | 1.15 |
| | containing 120° | 2.14 | 17.42 | 11.36 | 0.86 |
| | containing 150° | 2.58 | 18.65 | 12.01 | 1.10 |
| $WTe_2$ | {100} | 2.53 | 14.67 | 8.00 | 0.77 |
| | {010} | 2.85 | 16.89 | 12.85 | 1.28 |